\documentclass[twocolumn,twoside,slac_two]{revtex4}
\usepackage{graphicx}
\usepackage{fancyhdr}
\begin{document}

\title{The very early optical afterglow of GRBs, revealing  the nature 
of the ejecta}

%

\author{Y. Z. Fan}
\affiliation{Department of Physics, University of Nevada, Las Vegas
NV 89154, USA}
\affiliation{Purple Mountain Observatory, Chinese Academy of
Science, Nanjing 210008, China.}
\affiliation{National Astronomical Observatories, Chinese Academy of
Sciences, Beijing, 100012, China.}
\author{Bing Zhang}
\affiliation{Department of Physics, University of Nevada, Las Vegas
NV 89154, USA}
\author{D. M. Wei}
\affiliation{Purple Mountain Observatory, Chinese Academy of
Science, Nanjing 210008, China.}
\affiliation{National Astronomical Observatories, Chinese Academy of
Sciences, Beijing, 100012, China.}
\begin{abstract}
We show that if the GRB ejecta itself is magnetized or neutron-rich,
the very early afterglow of GRBs is very different from 
that powered by a pure ion fireball. In the {\it Swift} era, with the
well monitored early afterglow data, we can potentially diagnose
the ejecta composition and reveal the nature of the central GRB
engine. 

\end{abstract}

\maketitle

\thispagestyle{fancy}


\newcommand{\tot}{{\rm tot}}
\def\N{\nonumber}
\def\f{\frac}
\def\ga{\gamma}
\def\max{{\rm max}}
\def\min{{\rm min}}
\def\j{{\scriptscriptstyle (j)}}
\def\tot{{{\rm tot}}}
\def\sub{{{\rm sub}}}
\def\obs{{{\rm obs}}}
\def\dep{{{\rm dep}}}

\section{Introduction}
Since the discovery of the optical flash of GRB 990123 \cite{Akerlof1999},
great attention has been paid to the very early
optical afterglow (e.g., for theory see \cite{Sari1999, Meszaros1999,
Kobayashi2000, KS2000, CL2000, ZKM03, Li2003, Wu2003, KZ2003, NP2004};
for observation, see \cite{Akerlof2000, Kehoe00,  Fox03a, Fox03b,
Liw03, RES04}) of soft, long GRBs. The very early optical afterglow of
short GRBs and X-ray flashes have also been investigated in some
detail \cite{FZKM05, FWW04a}. 

In nearly all these theoretical papers, the fireballs are assumed to
be non-magnetized and  neutron-free. However, in principle, the ejecta
may be significantly magnetized (see \cite{LB03} and the references
therein) or contain a large amout of neutrons (e.g. \cite{DKK99,
B03}), or both (e.g. \cite{VPK03}). Recently we have studied the early
afterglow emission (especially the reverse shock emission) of a
magnetized ejecta and a neutron-rich ejecta
\cite{FWW04b,ZK05,FZW05}. Here we summarize  
the main results.  Comparing these results with the early afterglow
observations, we may get insight into the initial composition of the
outflow and reveal the nature of the central engine.  

\section{Reverse shock emission from magnetizated ejecta}
\subsection{Magnetized reverse shock emission from GRB 990123 and
GRB 021211} 

A bright optical flash was detected during the bright GRB 990123
(e.g., \cite{Akerlof1999}). The peak V-band flux was 8.9th magnitude.
After the peak, the flux drops as $t_{\rm obs}^{-2}$ (before the peak,
the flux increases as $t_{\rm obs}^{3.1}$). Such a sharp decrease has
been modeled by the emission from the shocked electrons contained in
the reverse shock region with adiabatic cooling \cite{Sari1999,
Meszaros1999}. With a more careful investigation, Fan et al.
found that by taking the reverse shock model and the physical
parameters found in modeling the multi-wavelength afterglow from the
forward shock emission, the theoretical peak flux of the optical flash
accounts for only $3\times 10^{-4}$ of the observed value ($\sim 1{\rm
Jy}$) \cite{FDHL02}. In order to remove this discrepancy, they
suggested that the electron and magnetic equipartition parameters,
$\epsilon_{\rm e}$ and $\epsilon_{\rm B}$, should be 0.61 and 0.39,
respectively. These are much different from the corresponding values
for the late afterglow ($\sim 0.1$ and $\sim 0.001$).   

In a more general discussion, Zhang, Kobayashi \& M\'{e}sz\'{a}ros
introduced a parameter $\cal{R}_{\rm B}$ (which is defined as
$(\epsilon_{\rm B}^{\rm r}/\epsilon_{\rm B}^{\rm f})^{1/2}$; where the
superscripts $r$ and $f$ represent the reverse shock region and
forward shock region, respectively) to trace the magnetization of the
GRB outflows \cite{ZKM03}. They found ${\cal{R}}_{\rm B}\sim$15 for
GRB 990123 and ${\cal{R}}_{\rm B}>1$ for GRB 021211.

These results have been also confirmed in several more detailed
analysis of both bursts \cite{KP03, PK04, MKP04}. As a result, it
is quite robust to say that the reverse shock region is magnetized, at
least for GRB 990123 and GRB 021211. In the following, we introduce a
parameter $\sigma$, the ratio between the electromagnetic energy flux
and the particle energy flux, to describe the initial magnetization
of the outflow (we assume that the magnetic field is ordered).  

\subsection{Reverse shock emission with mild magnetization}    
As the outflow interacts with the surrounding medium, two shocks
form. One is the forward shock expanding into the medium, and the
other is the reverse shock penetrating into the outflow. The forward
shock jump condition is the same as that of a pure hydrodynamical
fireball model \cite{BM76}, but the reverse shock jump condition is
different (see equations (2-5) of \cite{FWZ04} for the general form). 

\begin{figure}[t]
\centering
\includegraphics[width=9cm]{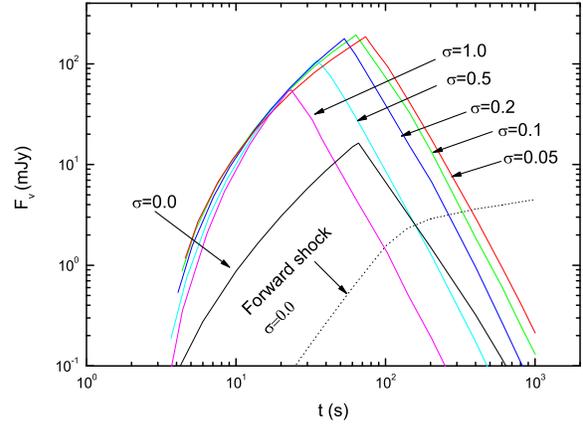}
\caption{
The very early R-band
($\nu_{\rm R}=4.6\times 10^{14}{\rm Hz}$) light curve powered by
the mildly magnetized outflow (the degrees of the magnetization
have been marked in the figure) interacting with the interstellar
medium. The parameters taken in the calculation are: $z=1$,
$E_{\rm kin}=10^{53}{\rm ergs}$, $L=2\times 10^{51}{\rm
ergs~s^{-1}}$, $p=2.2$, $\eta=300$, $n=1{\rm cm^{-3}}$, $\epsilon_{\rm
e}=0.3$ and the radiation efficiency $\epsilon=\epsilon_{\rm
e}$, where $E_{\rm kin}$ 
is the total initial energy of the outflow, $L$ is the luminosity of the 
$\gamma-$ray emission, $\eta$ is the initial bulk Lorentz factor
of the outflow, $p$ is the power-law distribution  index of the
shock-accelerated electrons, and
 $n$ is the number density of the medium. 
For $\sigma=0$ and the forward shock, it is assumed that
$\epsilon_{\rm B}=0.01$.
\ From ref~\cite{FWW04b}.
}
\label{fig_Lsigma_ISM}
\end{figure}

\begin{figure}[t]
\centering
\includegraphics[width=9cm]{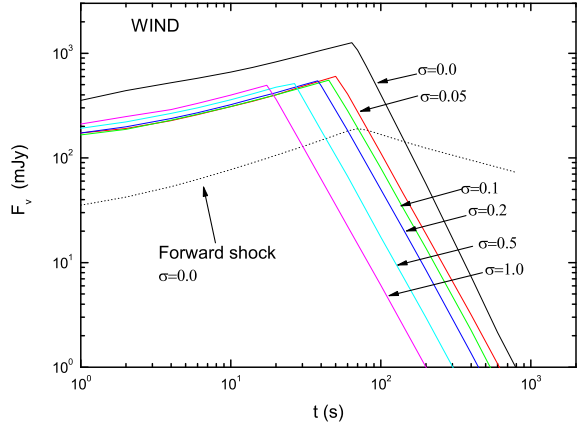}
\caption{The very early R-band
light curve powered by the mildly magnetized outflow (the degree
of the magnetization has been marked in the figure) interacting
with a stellar wind. The parameters taken here are the same to
those of figure 1 except $n=3\times 10^{35}R^{-2}{\rm
cm^{-3}}$. From ref~\cite{FWW04b}.
}
\label{fig_Lsigma_wind}
\end{figure}

The novel features include: {\bf 1}. At the reverse shock front, for
$\sigma>0.01$, the amplified ordered magnetic field ($B_{\rm ord}$) is
significantly stronger than that of the random one ($B_{\rm ran}$). In
the ISM medium, the typical synchrotron radiation $\nu_{\rm m}\propto
B_{\rm ord}$ is much lower than $\nu_{\rm R}\sim 4.6\times 10^{14}$Hz
and the reverse shock-accelerated electrons are 
in the slow cooling regime. The observed flux scales as
\begin{equation}
F_\nu\propto B_{\rm ord}\nu_{\rm m}^{(p-1)/2}\propto B_{\rm ord}^{(p+1)/2}.
\end{equation}
Therefore, the R-band reverse shock emission is stronger for a higher
$B_{\rm ord}$ (see Fig.\ref{fig_Lsigma_ISM}), which matches the
observation of GRB 990123 and GRB 021211. However, for $\sigma>0.1$,
the reverse shock is suppressed (in both \cite{FWW04b} and
\cite{ZK05}, the ideal MHD approximation is taken. If
magntic dissipation \cite{FWZ04} is taken into account, the
result would be different), so that the reverse shock peak level
starts to decrease (see Fig.\ref{fig_Lsigma_ISM}). 
In the wind medium case, the R-band reverse shock emission with
magnetization is weaker (see Fig. \ref{fig_Lsigma_wind}) since the
reverse shock is relativistic and the electrons are in the fast
cooling phase, and $\nu_{\rm m}$ is above $\nu_{\rm R}$. The observed
flux scales as
\begin{equation}
F_\nu\propto B_{\rm ord}\nu_{\rm c}^{1/2}\propto B_{\rm ord}^{-1/2},
\end{equation}
where $\nu_{\rm c}\propto B^{-3}$ is the cooling frequency. As a
result, $F_\nu$ decreaes with increasing $\sigma$. 

{\bf 2}. In the very early afterglow phase, the outflow
is ultrarelativistic. Due to the beaming effect, the area we view is
very narrow where the orientation of magnetic field is nearly the
same. The local high linear polarization can not be averaged
effectively. As a result, net high linear polarization is expected. 
By introducing a parameter $b=B_{\rm ord}/B_{\rm ran}$, the net linear
polarization can be approximated by (e.g., \cite{FWW04b})
\begin{equation}
\Pi_{\rm net}\approx 0.60{b^2\over 1+b^2}.
\end{equation}
Even for $\sigma\sim 0.01-0.1$, the linear polarization as high as
$30\%$ is expected.

\subsection{The reverse shock emission with arbitrary magnetization}

\begin{figure}[t]
\centering
\includegraphics[width=8cm]{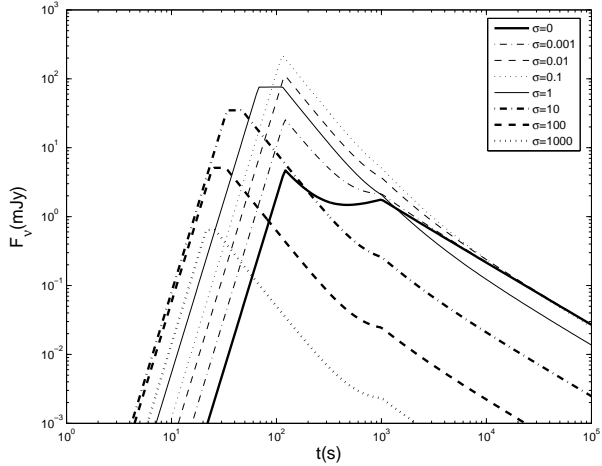}
\caption{
Sample early afterglow lightcurves for GRBs with an arbitrary
magnetization parameter $\sigma$. Following parameters are adopted.
$E_{\rm kin}=10^{52}$ergs, $T=20$s, $\eta=150$, $n=1{\rm cm^{-3}}$,
$\epsilon_{\rm e}^{\rm f}=\epsilon_{\rm e}^{\rm r}=0.1$, 
$\epsilon_{\rm B}^{\rm f}=0.001$, $p=2.2$, and $z=1$, where $T$ is the
duration of the GRB corrected by redshift. 
 Both the forward shock and the
reverse shock emission components are calculated and they are
superposed to get the final lightcurve. 
Lightcurves are calculated for
different $\sigma$ values.
 From ref~\cite{ZK05}.
}
\label{fig_Hsigma_ISM}
\end{figure}

Zhang \& Kobayashi have performed a detailed analytical investigation
on the reverse shock emission with arbitrary magnetization
\cite{ZK05}. For $\sigma<1$, i.e., the mildly magnetized regime, their
results are rather similar to that of \cite{FWW04b}. For $\sigma>1$,
i.e., the high-$\sigma$ regime, it is found out that the reverse shock
peak is broadened, mainly due to the separation of the shock crossing
radius and the deceleration radius of the outflow (see
Fig.\ref{fig_Hsigma_ISM}). This novel feature can be regarded as a
signature of high $\sigma$.

\begin{figure}[t]
\centering
\includegraphics[width=8cm]{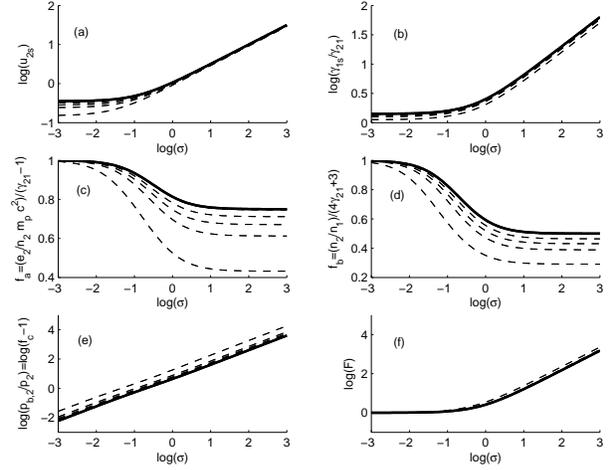}
\caption{
The variations of six parameters, i.e., $u_{2s}$,
$\gamma_{1s}/\gamma_{21}$, $e_2/n_2 m_p c^2$, $n_2/n_1$,
$p_{b,2}/p_2$, and $F$, as a function of $\sigma$. The thick solid
line is the Kennel-Coroniti solution \cite{KC84}, denoting a
$\gamma_{21} \gg 1$ regime. Here $u_{\rm 2s}$ is the radial four
velocity of region 2 measured in the shock frame; $\gamma_{\rm 1s}$ is
the bulk Lorentz factoe of region 1 measured in the shock frame;
$\gamma_{21}$ is the Lorentz factor of region 1 relative to region 2;
$e_2$ is the thermal energy density of region 2 (measured in the
comoving frame); $n_2$ ($n_1$) is the number density of region 2 (1)
measured in its comoving frame; $p_{b,2}$ ($p_2$) is the magnetic
field (thermal) pressure of region 2 measured in its comoving frame.
The dashed lines, starting from the one closest to the thick line, are
for $\gamma_{21}=1000, 100, 10, 5, 3, 1.5$, respectively. Again the
parameters $e_2 / n_2 m_p c^2$ and $n_2/n_1$ are normalized to
$(\gamma_{21}-1)$ and $(4\gamma_{21}+3)$, respectively.
\ From ref~\cite{ZK05}.
}\label{fig_Hsigma_supp}
\end{figure}

Another important result obtained is that the suppression factor of
the reverse shock in the strong magnetic field regime is only mild as
long as the shock is relativistic, and it saturates in the
high-$\sigma$ regime (see Fig.\ref{fig_Hsigma_supp}). This indicates
that strong relativistic shocks still exist in the high-$\sigma$
limit, which can effectively convert kinetic energy into heat. The
overall efficiency of converting jet energy into heat, however,
decreases with increasing $\sigma$, mainly because the fraction of the
kinetic energy in the total energy decreases \cite{ZK05}. These
results have been confirmed by Fan, Wei \& Zhang both analytically and
numerically \cite{FWZ04}, see their equations (17-18) and figures 1
and 2. 

\section{Very early optical afterglow lightcurves of neutron-fed GRBs} 

In the rest frame of the ejecta, the free neutron has a mean 
lifetime $\sim 900$s. The corresponding $\beta-$decay radius 
reads $R_{\beta}\sim 8\times 10^{15}\Gamma_{\rm n,2.5}$cm
\footnote{Through out the paper, the convention $Q_{\rm x}=Q/10^{\rm
x}$ is taken in cgs units.}, where $\Gamma_{\rm n}\sim 300$ is the
bulk Lorentz factor of the neutrons (below we call it as N-ejecta).
In the internal shock phase, the ion-ejecta (I-ejecta) has been
decelerated, but the neutrons are not. They move freely into the
medium and decay into protons, neutrinos and electrons. These dacay
products share their energy and momentum with the medium and form a
mixture (the trail)  moving with a bulk Lorentz factor about tens, the
actual velocity mainly depends on the density of the medium. 

Comparing with that of the pure fireball or the magnetized fireball, 
the early afterglow of neutron-fed GRBs is very complicated. The
detailed discussion has been presented in \cite{FZW05}. Here we only
summarize the main results.

\begin{figure}[t]
\centering
\includegraphics[width=8cm]{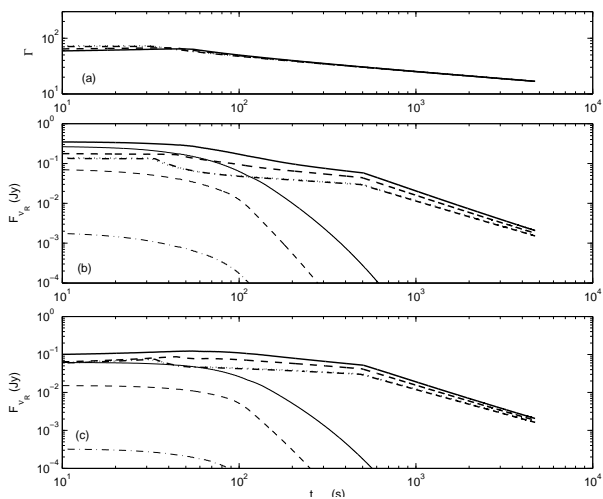}
\caption{
The early optical afterglow lightcurves of a neutron-fed long GRB in
the wind interaction case. (a) The dynamical evolution of the LF of
the shocked region as a function of time. (b) R-band lightcurves, with
the IC cooling effect due to the prompt $\gamma-$rays interacting with
the shocked regions being ignored. Thick lines include contributions
from all emission components, including the FS, RS and the neutron
decay trail. Thin lines are for trail emission only.  The dotted,
dash-dotted, dashed and solid lines represent
$\chi=0.0,~0.1,~0.5,~1.0$ respectively. Following input parameters are
adopted in the calculations: $E_{\rm tot}=2.0\times 10^{53}{\rm
ergs}$, $\Delta=10^{12}$cm, $z=1$ [i.e. $d_{\rm L}=2.2\times
10^{28}{\rm cm}$, which corresponds to the standard
$(\Omega_m,\Omega_\Lambda)=(0.3,0.7)$ $\Lambda$CDM cosmological
model], $\Gamma_{\rm n}=300$, $\Gamma_{\rm m}=200$, $\Gamma_{\rm
s,n}=30$, and $n=10^{35}{\rm cm^{-3}}R^{-2}$ (i.e. $A_*=1/3$),
respectively. The parameters $\epsilon_{\rm e}=0.1$, $\epsilon_{\rm
B}=0.01$ and $p=2.3$ are adopted for the FS and RS shocks as well as
the trail. Where $E_{\rm tot}$ is the total energy of the initial
outflow; $\Delta$ is the width of the initial I-ejecta; $\Gamma_{\rm
n}$ ($\Gamma_{\rm s,n}$) is the bulk Lorentz factor of fast (slow)
neutrons; $\Gamma_{\rm m}$ is the initial bulk Lorentz factor of
I-ejecta.  (c) R-band lightcurves, but the IC cooling effect due to
the prompt $\gamma-$rays overlaping with the shocked region and the
trail has been taken into account. The averaged $\gamma-$ray
luminosity is taken as $L_\gamma=10^{51}{\rm ergs}~{\rm
s}^{-1}$. Other paramters and line styles are the same as those in
(b).  From ref~\cite{FZW05}.
}
\label{fig_Neutron_wind}
\end{figure}

If the medium is a pre-stellar wind, the neutron
trail moves slowly, mainly because the medium inertia is too
large. The trail and the I-ejecta do not separate from each other,
and a forward shock propagates into the trail directly. Three
components contribute to the final emission, i.e. the forward shock,
the reverse shock propagates into the I-ejecta, and the unshocked
trail emission. The latter is significant when $\chi$, 
the ratio of neutrons to protons, is large, since
the internal energy of the unshocked trail is large when the medium
density is high. A typical neutron-rich wind-interaction lightcurve
is a characterized by a prominent early plateau lasting for $\sim 100$
s followed by a normal power-law decay (Fig.\ref{fig_Neutron_wind}). 
We also show that in the wind case, the IC cooling effect due to the
overlaping of the initial prompt $\gamma-$ray with the shocks and the
trail suppresses the
very early R-band afterglow significantly. The neutron-fed
signature is also dimmed  (see Fig.\ref{fig_Neutron_wind}(b) and
Fig.\ref{fig_Neutron_wind}(c) for a comparison).

\begin{figure}[t]
\centering
\includegraphics[width=8cm]{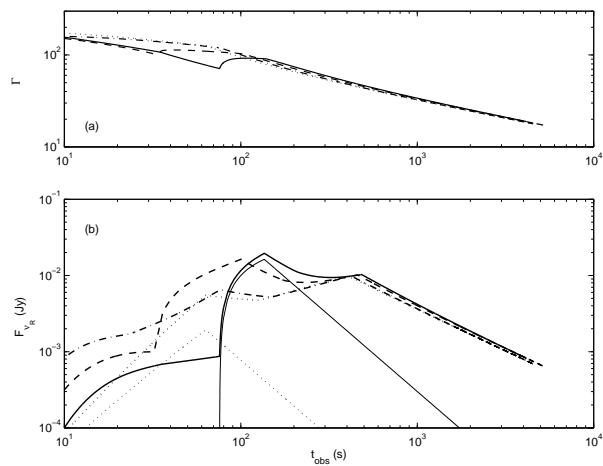}
\caption{
The early opticcal afterglow lightcurves of a neutron-fed long GRB in
the ISM interaction case. (a) The dynamical evolution
of the region shocked by FS as a function of time. The dotted,
dash-dotted, dashed, and solid lines represent
$\chi=0.0,~0.1,~0.5,~1.0$ respectively. (b) R-band lightcurves. Line
styles are the same as in (a). Thick lines represent the total early
R-band lightcurves, while the thin lines are for RS emission only. 
Only the RS emission for $\chi=0$ and the RRS emission for $\chi=1$
cases are plotted.
The initial parameters are the same as those listed in the caption of
Figure \ref{fig_Neutron_wind}, except that $n=1{\rm cm^{-3}}$ is adopted.
 From ref~\cite{FZW05}.
}
\label{fig_Neutron_ISM}
\end{figure}

If the medium is a constant density ISM, part of
the neutron decay products fall onto the medium, and the trail moves
faster than the I-ejecta (In \cite{FZW05}, it is assumed that the I-ejecta 
moves slower than the fast neutrons). A gap likely forms between the
leading trail 
and the I-ejecta. The former forms a distinct trail ejecta (T-ejecta)
which interacts with the out trail or ISM. The latter catches up later
and gives rise 
to a rebrigtening signature. Before collision, the radiation is
dominated by the forward shock emission. During the collision, both
the forward shock emission and the refreshed shocks (especially the
refreshed reverse shock) are important. The unshocked trail emission
is not important in this case. A typical neutron-rich ISM-interaction 
lightcurve is characterized by a slow initial rising lightcurve
followed by a prominent bump signature around tens to hundreds of seconds
(Fig.\ref{fig_Neutron_ISM}). 

For all the cases, the predicted signatures can be detected by the
UVOT on board the {\em Swift} observatory. However, most of these
signatures (such as the plateau and the bump signature) are not
exclusively for neutron decay. More detailed modeling and case study are
needed to verify the existence of the neutron component.

\section{Discussion}
Most of the current afterglow observations take place hours after the
burst trigger. At this stage, the observed afterglow emission are
powered by the forward shock, so that essentially all the initial
information of the ejecta is lost. In order to diagnose the ejecta
composition, well-monitored early afterglow data are needed, since the
early afterglow emission, especially in the optical band, is believed
to be dominated by the reverse shock emission and possibly the trail
emission. The reverse shock propagates into the ejecta, so the
emission property, especially the linear polarization degree, depends
on the magnetization of the ejecta. The neutron signature emission
lasts only tens to hundreds of seconds, which overlaps with the very
early shocks emission. Only rapid responce to the trigger can
catch it. 

In this proceedings paper, we have summarized the early afterglow
signatures of magnetized GRBs and neutron-rich GRBs. These
signatures are likely detectable by the Ultraviolet Optical Telecope
(UVOT) on board the {\em Swift} observatory. Close monitoring of early
afterglows from 10s to 1000s of seconds, when combined with detailed
theoretical modeling, could be used to diagnose their existence,
which may in turn help us to reveal  the nature of the  GRB central
engine source. 

Currently, there are just two well studied cases. i.e. GRB 990123,
and GRB 021211, whose early afterglow lightcurves are well consistent
with reverse shock emission from a moderately-magnetized flow
\cite{FDHL02,ZKM03,KP03,PK04,MKP04}. However, the magnetization
itself does not mean the outflow is initially magnetized since in the
internal shock phase, random magnetic fields are generated, significant
part of which can not be dissipated effectively. These magnetic fields
could be retained in the external shock phase to dominate the
reverse shock synchrotron emission. In this case, there is no net
polarization expected. Consequently, the early polarization detection
is necessary to draw definitive conclusions on the initial
magnetization of the ejecta.

\bigskip 
\begin{acknowledgments}
Y.Z.F thanks the conference organizers for  partial support.
This work is supported by NASA
NNG04GD51G and a NASA Swift GI (Cycle 1) program (for B.Z.),
the National Natural Science Foundation (grants 10225314 and 10233010)
of China, and the National 973 Project on 
Fundamental Researches of China (NKBRSF G19990754) (for D.M.W.).
\end{acknowledgments}

\bigskip 

\end{document}